\documentclass[twocolumn,showpacs,preprintnumbers,amsmath,amssymb,prl]{revtex4}
\usepackage{graphicx}
\begin{document}
\title{Orthogonality \\of Biphoton Polarization States}
\author{ M.~V.~Chekhova, L.~A.~Krivitsky, S.~P.~Kulik, and G.A.Maslennikov}
\address{Department of Physics, Moscow M.V. Lomonosov State
University, 119992 Moscow, Russia} \widetext\vspace*{-10mm}

\begin{abstract}
\begin{center}\parbox{14.5cm}
{Orthogonality of two-photon polarization states belonging to a
single frequency and spatial mode is demonstrated experimentally,
in a generalization of the well-known anti-correlation 'dip'
experiment.}
\end{center}
\end{abstract}

\maketitle
\narrowtext \vspace{-10mm}

Orthogonality, one of the basic mathematical concepts, plays an
important role in physics, especially quantum physics and, in
particular, quantum optics. A well-known example including both
classical and quantum cases is orthogonality of two polarization
modes of electromagnetic radiation.  Physically, orthogonality of
two arbitrary polarization states means that if light is prepared
in a certain (in the general case, elliptic) polarization state it
will not pass through a filter selecting the orthogonal state. (A
filter selecting an arbitrary polarization state can be made of a
rotatable quarter-wave plate and a rotatable linear polarization
filter~\cite{Shurkliff}.) Orthogonality of polarization states has
an explicit representation on the Poincar\'{e} sphere where each
polarization state is depicted by a point. The state orthogonal to
a given one is shown by a point placed on the opposite side of the
same diameter. Examples are states with vertical and horizontal
linear polarization, right- and left- circularly polarized states,
and any two elliptically polarized states with opposite directions
of rotation and inverse axis ratios. This concept of orthogonality
relates to both classical polarization states of light and
single-photon quantum states of polarized light~\cite{kljetp}.
Mathematically, orthogonality of two polarization states means
that the scalar product of two corresponding Jones
vectors~\cite{Shurkliff} is equal to zero. In quantum optics, this
corresponds to zero scalar product of polarization state vectors,
for instance, state vectors of single-photon states. This is a
particular case of the general rule: orthogonality of quantum
states means that their scalar product is equal to zero.

However, in addition to single-photon states there are other types
of nonclassical light. In quantum optics, one of the central roles
is played by two-photon states, which are most easily generated
via spontaneous parametric down-conversion (SPDC)~\cite{photons}.
In a two-photon state, radiation consists of photon pairs, often
called biphotons, that are correlated in frequency, wavevector,
moment of birth, and polarization. Focusing on the case of
collinear frequency-degenerate SPDC, here we will discuss the
so-called single-mode biphotons. Although even in the
frequency-degenerate collinear case SPDC has a finite frequency
and angular spectrum, under certain experimental conditions such
biphotons can be treated as relating to a single frequency and
angular mode.

One can show that the polarization state of a single-mode
biphoton~\cite{burkl} can be described as a qutrit, a three- state
quantum system. Using qutrits instead of qubits for the
transmission of quantum information has been previously
discussed~\cite{trits}, in particular, in connection with ternary
cryptography
protocols~\cite{bechmann},~\cite{bruss},~\cite{peres}. The general
case of a qutrit represented by a biphoton in an arbitrary pure
polarization state is given by the state vector

\begin{equation}
|\psi\rangle=c_1|2,0\rangle+c_2|1,1\rangle+c_3|0,2\rangle,
\label{1}
\end{equation}
where $c_i$ are complex amplitudes satisfying the normalization
condition, $|c_1|^2+|c_2|^2+|c_3|^2=1$, and $|n,m\rangle$ denotes
a two-photon ($n+m=2$) state with $n$ photons polarized
horizontally and $m$ photons polarized
vertically~\cite{indistinguishable}. In~\cite{quarkobzor}, it is
shown that the state (1) allows an explicit representation on the
Poincar\'{e} sphere. It can be written in the form of two
arbitrarily polarized correlated single-photon states,

\begin{equation}
|\psi\rangle=\frac{a^{\dagger}(\theta,\phi)a^{\dagger}(\theta',\phi')
|\rm{vac}\rangle}{||a^{\dagger}(\theta,\phi)a^{\dagger}(\theta',\phi')
|\rm{vac}\rangle||}. \label{2}
\end{equation}

Here, $a^{\dagger}(\theta, \phi)$ and $a^{\dagger}(\theta',
\phi')$ are operators of photon creation in arbitrary
polarization modes given by the coordinates $\theta, \phi$,
$\theta', \phi'$ on the Poincar\'{e} sphere:

\begin{equation}
a^{\dagger}(\theta, \phi)={\rm cos}(\theta/2)a_H^{\dagger}+{\rm
e}^{i\phi}{\rm sin}(\theta/2)a_V^{\dagger},
 \label{3}
\end{equation}
where $a_{H,V}^{\dagger}$ are photon creation operators in the
horizontal and vertical linear polarization modes, and similarly
for $a^{\dagger}(\theta',\phi')$. Axial angles $\theta,\theta'$
and azimuthal angles $\phi,\phi'$~\cite{angles} are in one-to-one
correspondence with the four parameters describing the state (1),
which are, for instance, $d_1=|c_1|$, $d_3=|c_3|$, $\phi_1={\rm
arg}(c_1)-{\rm arg}(c_2)$, $\phi_3={\rm arg}(c_3)-{\rm
arg}(c_2)$. Representation (2) means that a biphoton of arbitrary
polarization can be shown as a pair of points on the Poincar\'{e}
sphere. It turns out that the Stokes vector of a biphoton is
simply a normalized sum of the Stokes vectors of photons forming
it ('biphoton halves'), and the polarization degree $P$ of the
pair is given by the angle $\sigma$ at which the pair can be seen
from the sphere center.

A question arises: what does orthogonality of two biphoton
polarization states mean? As usually in quantum mechanics, it
means that the product of their state vectors is equal to zero.
For instance, orthogonality of two biphotons $\psi_{ab}\equiv
\frac{a^{\dagger}b^{\dagger}|{\rm vac}\rangle}{||a^{\dagger}
b^{\dagger}|{\rm vac}\rangle||}$ and $\psi_{cd}\equiv
\frac{c^{\dagger}d^{\dagger}|{\rm vac}\rangle}{||c^{\dagger}
d^{\dagger}|{\rm vac}\rangle||}$, with the operators of photon
creation in arbitrary polarization modes denoted now by
$a^{\dagger},b^{\dagger}, c^{\dagger},d^{\dagger}$, means that

\begin{equation}
\langle {\rm vac}| c d a^{\dagger} b^{\dagger}| {\rm
vac}\rangle=0. \label{4}
\end{equation}

What does orthogonality of two biphotons mean from the viewpoint
of physics? This question was answered in~\cite{orthogonality}
where an operational criterion of orthogonality for arbitrarily
polarized biphotons was formulated. Namely, orthogonality of two
biphotons can be tested using a simple setup consisting of a
non-polarizing beamsplitter, two detectors installed in its two
output ports, with an arbitrary polarization filter inserted at
the input of each detector, and a coincidence circuit. A biphoton
is registered if there is a coincidence of photocounts from the
two detectors. Let a biphoton $|\psi_{ab}\rangle$ be at the input
and the filters in the output ports of the beamsplitter select
polarization states corresponding to photon creation operators
$c^{\dagger}, d^{\dagger}$. Then orthogonality of
$|\psi_{ab}\rangle$ and $|\psi_{cd}\rangle$ is equivalent to the
absence of coincidences~\cite{absence} in such a setup. Note that
orthogonality of any two polarization states among
$a^{\dagger}{\rm |vac\rangle}$, $b^{\dagger}{\rm |vac\rangle}$,
$c^{\dagger}{\rm |vac\rangle}$, $d^{\dagger}{\rm |vac\rangle}$ is
not required.

Such an experiment is the most general case of the well-known
anticorrelation experiment~\cite{hong}. Earlier, a particular case
of this polarization version of anticorrelation experiment has
been performed for type-II SPDC~\cite{typeiianti}. The absence of
coincidences in the anticorrelation experiment~\cite{typeiianti}
can be interpreted as orthogonality of the states
$|HV\rangle\equiv a_H^{\dagger}a_V^{\dagger}|{\rm vac}\rangle$ and
$|D\overline{D}\rangle\equiv
a_{45}^{\dagger}a_{-45}^{\dagger}|{\rm vac}\rangle$, a pair of
photons polarized linearly at angles $\pm 45^{\circ}$ to the
vertical axis. Similarly, in~\cite{ternary}, orthogonality of the
states $|HV\rangle$ and $|D\overline{D}\rangle$ to the state of
right- and left polarized photons, $|RL\rangle$, has been
demonstrated. In both cases, orthogonal biphotons had zero
polarization degree, which means that they were pairs of
orthogonally polarized photons. Another example of a basis formed
by three mutually orthogonal biphotons with zero polarization
degree was demonstrated in~\cite{trifonov}.

Several examples of orthogonal biphotons of other polarization
degrees are shown in Fig.1. Fig.1a shows three mutually orthogonal
states $|HV\rangle$, $|D\overline{D}\rangle$, $|RL\rangle$ of
biphotons with zero polarization degree studied
in~\cite{typeiianti} and~\cite{ternary}. All three states are
biphotons consisting of two orthogonal photons; at the same time,
no photon forming a biphoton is orthogonal to any photon of the
other two biphotons. Three biphotons shown in Fig.1a form an
orthogonal basis.

Whenever a biphoton is fixed (two points are fixed on the
Poincar\'{e} sphere), there are infinitely many biphotons
orthogonal to it. Replacing the Poincar\'{e} sphere by the globe,
one can pick two spots denoting a biphoton to be, for instance,
Moscow (Russia) and Turin (Italy). Then, one should make a choice
for the third point. Let it be Baltimore(MD, USA). Then the fourth
point is found from Eq.~(4), and it turns out to be near New
Zealand and the Bounty isles. So, the biphoton 'Moscow - Turin' is
orthogonal to the biphoton 'Baltimore - Bounty'!

The idea of our experiment was to prepare some arbitrary input
biphoton state, to make the registration part select the state
orthogonal to it, and to demonstrate orthogonality by scanning the
parameters of the input and registered states around the minimum
of coincidence counting rate. We chose the input biphoton state to
be a pair of photons polarized linearly at the opposite angles to
the horizontal direction, the polarization degree being $P=0.5$.
On the Poincar\'{e} sphere, this is shown as two points on the
equator placed symmetrically at the angles $\pm 74.5^{\circ}$ with
respect to the 'H' axis, which corresponds to photons polarized
linearly at the angles $37.25^{\circ}$ to the horizontal axis. For
the reasons that we will explain later, it is convenient to make
one of the polarization filters in the registration part select
the state polarized linearly at the angle $45^{\circ}$ to the
horizontal axis. The other filter, as one can easily find, should
then select linear polarization at the angle $60^{\circ}$ to the
horizontal axis. This configuration, which is used in one of our
experiments, is shown in Fig.1b. Finally, Fig.1c gives an example
of two orthogonal biphotons with polarization degree $P=0.5$ in a
'non-plane' configuration. This time, the points denoting the
biphoton 'ab' are placed on the 'Greenwich meridian', if we follow
the globe terminology. Again, one of the polarizers selects the
state polarized linearly at $45^{\circ}$. However, the position of
the other polarizer is changed: now, it should select linear
polarization at the angle $-60^{\circ}$ to the horizontal axis.
Under each Poincar\'{e} sphere in Fig.1, the corresponding
polarization states are shown schematically. In all these
examples, there is no orthogonality between separate photons, or
'halves of biphotons'.

The experimental setup is shown in Fig.2. Collinear
frequency-degenerate SPDC is generated in two similar type-I
lithium iodate crystals of length 1 cm, cw radiation of argon
laser at wavelength 351 nm used as the pump. The optic axis of the
first crystal is in the vertical plane while the optic axis of the
other crystal is in the horizontal plane. The two-photon state
generated after the crystals is of the form (1), with $c_2=0$, the
amplitudes $d_1$ and $d_3$ can be varied by rotating the
$\lambda/2$ plate in the pump beam, and the phase
$\Delta\phi\equiv\phi_3-\phi_1$ can be varied by tilting the two
quartz plates QP, whose optic axes are oriented in the vertical
plane. The pump after the crystals is cut off by a UV mirror UVM.
Spatial and frequency filtering of the SPDC radiation is performed
by a pinhole P and an interference filter IF with $702$ nm central
wavelength and $3$ nm bandwidth. The right-hand side of the setup
shows the registration part (the Brown-Twiss interferometer). It
includes a non-polarizing beamsplitter BS and two detectors
(photomultiplier tubes) D1, D2 inserted in its output ports. At
the input of each detector, there is a polarization filter
consisting of a rotatable quarter-wave plate QWP1,2~\cite{qp} and
a rotatable polarizer P1,2. Coincidences between the photocounts
of the detectors are registered using a coincidence circuit with a
resolution $T_c=5.5$ ns.

The measurements were performed for two configurations shown in
Figs 1b,c. The plates QP were tilted so that the phase
$\Delta\phi$ was equal to $\pi$. Then the angle $\chi$ of the
$\lambda/2$ plate was scanned from $0^{\circ}$ to $90^{\circ}$. As
a result, the two points corresponding to the produced biphoton
state travelled on the Poincar\'e sphere: first, from the 'VV'
point ($\chi=0^{\circ}$) to the 'HH' point ($\chi=45^{\circ}$)
symmetrically along the opposite sides of the equator, then again
to the 'VV' point ($\chi=90^{\circ}$) but this time, along the
opposite sides of the 'Greenwich meridian'. This way, both cases
shown in Fig.1b,c were realized. In the first run, the polarizer
P1,2 orientations were fixed as in Fig.1b: $\zeta_1=45^{\circ}$
and $\zeta_2=60^{\circ}$. In the dependence of coincidence
counting rate on $\chi$ (Fig.3a), the minimum was at
$\chi=30^{\circ}$, which corresponded to $d_1^2/d_3^2=3$. For this
point, the ratio $d_1^2/d_3^2$ was measured using the tomography
procedure developed in~\cite{tomography}; this ratio turned out to
be $3.4\pm 0.8$. The $45^{\circ}$ orientation of P1 is convenient
since in this case, rotation of the half-wave plate in the pump
beam does not lead to the variation of D1 single counting rate
$R_1$.

In the next run, the orientations of P1, P2 were fixed as in
Fig.2c: $\zeta_1=45^{\circ}$ and $\zeta_2=-60^{\circ}$. In this
case, the minimum was achieved for $\chi=60^{\circ}$ (Fig.3b),
corresponding to $d_1^2/d_3^2=1/3$.

Finally, orthogonality of two biphoton states was checked by
fixing all parameters ($\chi$, $\zeta_{1,2}$, $\Delta\phi$) in the
configuration shown in Fig.1b and then scanning them around their
optimal values. The plot in Fig.4 shows the dependence of the
coincidence counting rate on the orientation of polarizer P1, with
the other polarizer fixed at $60^{\circ}$ and the half-wave plate
in the pump beam fixed at $\chi=30^{\circ}$. Similar dependencies
were obtained for scanning $\zeta_2$ and $\Delta\phi$.

Two important notes should be made about the measurement
procedure. First, when the angle $\chi$ is scanned with fixed
positions of the two polarizers (Fig.3a), a considerable
modulation in the singles counting rate $R_2$ of detector D2 is
observed (Fig.5). This is quite natural since in the course of
$\chi$ variation the state of biphoton light changes from being
vertically polarized through completely non-polarized state (at
$\chi=22.5^{\circ}$) to being horizontally polarized. With the
polarizer P1 oriented at $\pm60^{\circ}$, the intensity of
transmitted light should vary three times, as it indeed does in
Fig.5. However, the minimum of the coincidence counting rate $R_c$
does not correspond to the point where $R_2$ is minimal; according
to the calculation, it occurs at $\chi=30^{\circ}$. The second
remark is that in all kinds of anticorrelation experiments, the
coincidence counting rate in the minimum is given by the level of
accidental coincidence counting rate, which corresponds to the
normalized second-order Glauber's correlation function
$g^{(2)}=1$. To demonstrate this, instead of the coincidence
counting rate, in all plots we present $g^{(2)}$~\cite{g2} instead
of $R_c$.

To find the dependence of the coincidence counting rate $R_c$ on
all parameters, one can write

\begin{equation}
R_c\sim|\langle {\rm vac}| c d a^{\dagger} b^{\dagger}| {\rm
vac}\rangle|^2. \label{5}
\end{equation}

The operators $a^{\dagger}$, $b^{\dagger}$, $c$, $d$ are
substituted in (5) in the form (3), with the angles $\phi$ taking
the values $0,\pi,0,0$, respectively. The axial angles for $a,b$
can be expressed as functions of $\chi$ via the relations given
in~\cite{orthogonality} and the formulas $d_1={\rm sin}(2\chi)$,
$d_3={\rm cos}(2\chi)$. The axial angles for $c,d$ can be
calculated as functions of $\zeta_{1,2}$, the angles of P1,P2
orientations. Then we obtain the coincidence counting rate,

\begin{equation}
R_c\sim\left[{\rm cos}\zeta_1{\rm cos}\zeta_2{\rm
sin}(2\chi)-{\rm sin}\zeta_1{\rm sin}\zeta_2{\rm
cos}(2\chi)\right]^2, \label{6}
\end{equation}
and the single-photon counting rate in detector 2,

\begin{equation}
R_2\sim{\rm cos}^2\zeta_2{\rm sin}^2(2\chi)+{\rm
sin}^2\zeta_2{\rm cos}^2(2\chi). \label{7}
\end{equation}

Equations (6,7) were used to plot the theoretical dependencies
shown in Figs.3,4.

To conclude, we have experimentally demonstrated orthogonality of
two biphotons having polarization degree between $0$ and $1$. Our
experiment is a generalization of the 'anti-correlation dip'
experiment to the case of arbitrarily polarized photon pairs. The
observed effect can find applications in ternary quantum
cryptography protocols~\cite{cryptography}.

This work was supported in part by the Russian Foundation for
Basic Research grants \#02-02-16664,\#03-02-16444, INTAS grant
\#01-2122, and the Russian program of scientific school support
(\#166.2003.02)

\begin{figure}
\includegraphics[width=1\columnwidth]{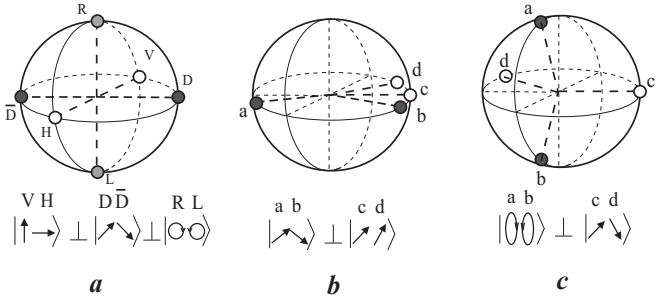}
\caption{Different cases of orthogonal biphotons: a - three
orthogonal non-polarized biphotons, $|HV\rangle$, $|RL\rangle$,
and $|D\overline{D}\rangle$; b - two orthogonal biphotons formed
by linearly polarized photons. The input biphoton has polarization
degree $P=0.5$ and the polarizers are at $45^{\circ}$ and
$60^{\circ}$ to the horizontal axis. c - the 'non-plane' version:
the input state also has $P=0.5$ but is formed by elliptically
polarized photons. }\label{States}
\end{figure}
\begin{figure}
\includegraphics[width=1\columnwidth]{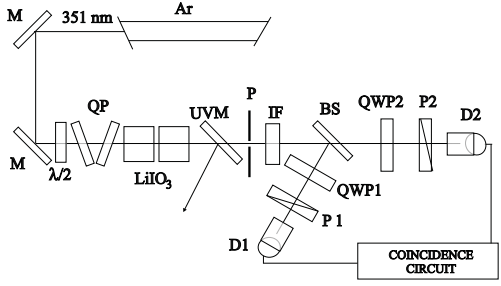}
\caption{The experimental setup. SPDC is excited in two type-I
$\hbox{LiIO}_3$ crystals with the optic axes in orthogonal planes;
the first crystal generates $|2,0\rangle$ and the second one,
$|0,2\rangle$. Quartz plates QP enable variation of the phase
between the two states from $0$ to $\pi$. UVM is a UV mirror, P a
pinhole, IF an interference filter, BS a nonpolarizing
beamsplitter, QWP1,2 are quarter-wave plates, P1,2 rotatable
polarizers, D1,2 detectors.}\label{Scheme}
\end{figure}
\begin{figure}
\includegraphics[width=0.7\columnwidth]{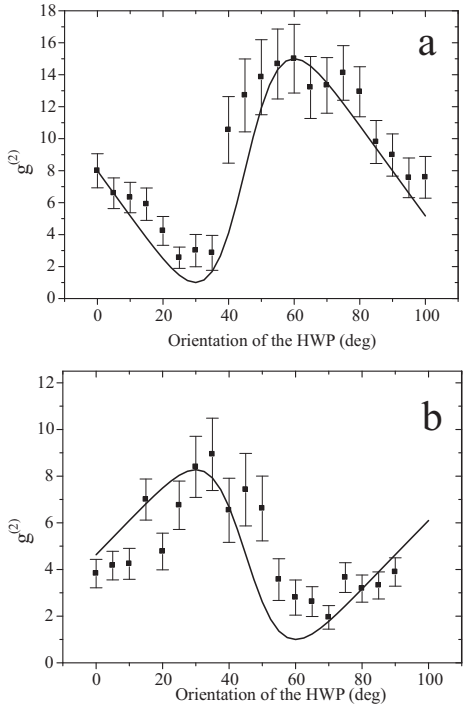}
\caption{Normalized second-order correlation function versus the
angle $\chi$ for $\Delta\phi=\pi$ and the polarizers P1,2 fixed at
$45^{\circ}$ and $60^{\circ}$, respectively (a) and at
$45^{\circ}$ and $-60^{\circ}$, respectively (b)}
\end{figure}
\begin{figure}
\includegraphics[width=0.7\columnwidth]{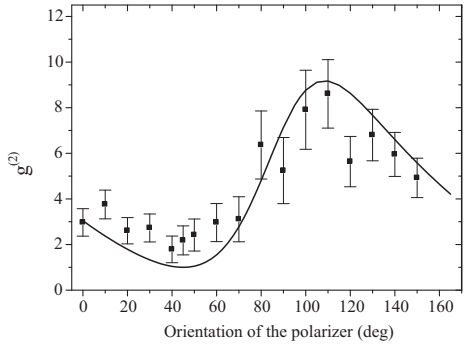}
\caption{Normalized second-order correlation function versus the
angle of the polarizer P1 orientation with the half-wave plate
fixed at $\chi=30^{\circ}$ and the phase
$\Delta\phi=\pi$}\label{polarizer}
\end{figure}
\begin{figure}
\includegraphics[width=0.7\columnwidth]{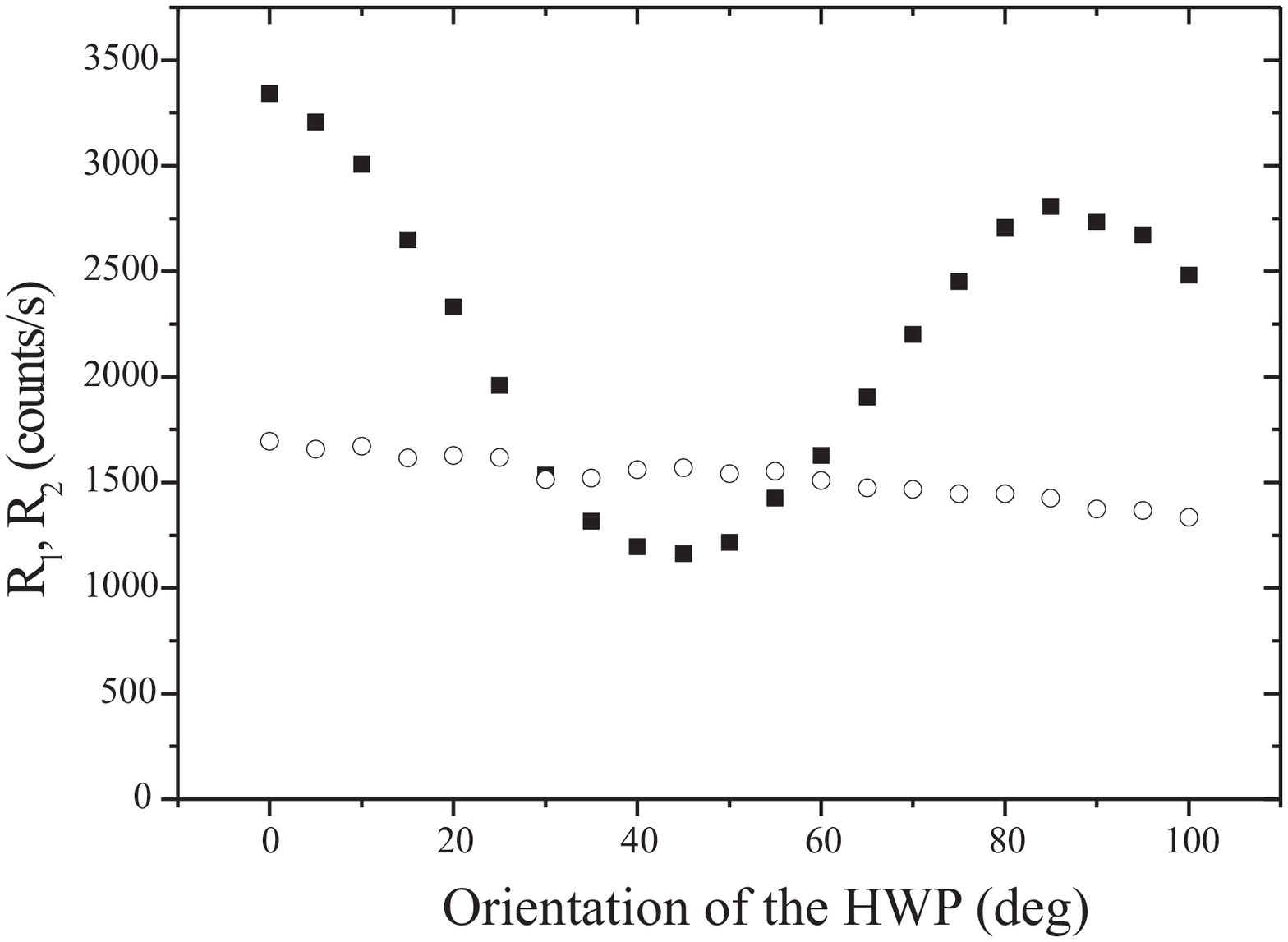}
\caption{Single-photon counting rates at detectors D1 (circles)
and D2 (squares) versus the angle $\chi$ for $\Delta\phi=\pi$ and
the polarizers P1,2 fixed at $45^{\circ}$ and $60^{\circ}$,
respectively. The total decrease in the counting rates is caused
by a gradual decrease in the pump power.}\label{singles}
\end{figure}


\begin{thebibliography}{1}
\bibitem{indistinguishable}Here, it is important that the two
photons may differ only in polarization. In the case of
distinguishable photons, the state $|1,1\rangle$ is replaced by
two Bell states $\Psi^{\pm}$.
\bibitem{angles}$\theta$ is measured with
respect to the 'H' axis, giving the state of horizontal
polarization and $\phi$ is measured with respect to the 'D' axis
giving  $45^{\circ}$ linear polarization.
\bibitem{absence}Coincidences are not completely
absent but their rate corresponds to the Poissonian (accidental)
level.
\bibitem{qp}Although quarter-wave plates are necessary in the general case, in the
measurements performed here they could be omitted.
\end{thebibliography}
\end{document}